\documentclass[rotate,psfig,epsf]{article}
\usepackage{psfig,amsmath}
\newcommand {\apgt} {\ {\raise-.5ex\hbox{$\buildrel>\over\sim$}}\ }
\newcommand {\aplt} {\ {\raise-.5ex\hbox{$\buildrel<\over\sim$}}\ } 
\title{Analysis of RXTE-PCA Observations of SMC X-1}
\author{S.\c{C}. \.{I}nam$^1$, A. Baykal$^2$, E. Beklen$^{2,3}$ \\ $^1$ Department of Electrical and Electronics Engineering,\\  Ba\c{s}kent University, 06530 Ankara, Turkey \\ inam$@$baskent.edu.tr \\
$^2$ Physics Department,\\ Middle East Technical University, 06531 Ankara, Turkey \\ elif$@$astroa.physics.metu.edu.tr \\ 
$^3$ Physics Department, S\"{u}leyman Demirel University, 32260 Isparta, Turkey
}
\date{}

\begin{document}

\maketitle
\begin{abstract}

We present timing and spectral analysis of RXTE-PCA observations of SMC X-1 between January 1996 and December 2003. From observations around 30 August 1996 with a time span of $\sim 6$ days, we obtain a precise timing solution for the source and resolve the eccentricity as 0.00089(6). We find an orbital decay rate of $\dot P_{orb}/P_{orb} =-3.402(7) \times 10^{-6}$ yr$^{-1}$ which is close to the previous results. Using our timing analysis and the previous studies, we construct a $\sim 30$ year long pulse period history of the source. We show that frequency derivative shows long (i.e. more than a few years) and short (i.e. order of days) term fluctuations. From the spectral analysis, we found that all spectral parameters except Hydrogen column density showed no significant variation with time and X-ray flux. Hydrogen column density is found to be higher as X-ray flux gets lower. This may be due to the increase in soft absorption when the pulsar is partially obscured as in Her X-1 or may just be an artifact of the tail of a soft excess in energy spectrum. 

{\bf{Keywords:}} X-rays: binaries, pulsars,  individual: SMC X-1 , stars: neutron, accretion, accretion discs 
\end{abstract}
\section{Introduction}
The high mass X-ray binary (HMXB) system SMC X-1/Sk 160 consists of a neutron star with a mass of $\sim1.06M_{\odot}$ (van der Meer et al. 2007) and a spin period of 0.71s (Lucke et al.1976), accreting from the B0I star Sk 160 with a mass of  $\sim17.2M_{\odot}$ (Reynolds et al. 1993). Orbital period of the system is $\sim3.9$ days (Schreier et al. 1972). The system has been observed with several observatories since it was discovered during a rocket flight (Price et al. 1971). It is the only discovered HMXB with a supergiant companion in SMC so far (Galache et al. 2008).

From the pulse timing studies, the source was observed to be spinning up since it was discovered (Wojdowski et al. 1998). Levine et al. (1993) and Wojdowski et al (1998) found an orbital period decay in the system with $P_{orb}/\dot{P}_{orb}$ of $\sim 3.4\times 10^{-6}$yr$^{-1}$.  

SMC X-1 also exhibits super-orbital X-ray flux variations like the 35 day cycle of Her X-1 (Gruber $\&$ Rothschild, 1984). Average super-orbital period of the source is $\sim$ 55 days (Wojdowski et al. 1998; Trowbridge et al. 2007). The super-orbital X-ray flux variations of SMC X-1 are thought to be due to the precession of a warped accretion disk (Wojdowski et al. 1998).

In this paper, we present timing and spectral analysis of RXTE (Rossi X-ray Timing Explorer) observations of SMC X-1 between January 1996 and December 2003. In the next section, we give brief information about instruments and observations. In Section 3, we present the results of timing analysis, including pulse period history and timing solution of the source. In Section 4, X-ray spectral analysis of the source is presented. 

\section{Instruments and Observations}

We analyzed data from Proportional Counter Array (PCA) onboard RXTE (Jahoda et al 1996) of SMC X-1 between January 1996 and December 2003 (MJD 50093-52988). 

The RXTE-PCA consists of an array of 5 Proportional Counters (PCU) operating in the 2-60 keV energy range, with a total effective area of approximately 7000 cm$^2$ and a field of view of $\sim$1$^{\circ}$ FWHM.
Data obtained from top xenon layers (L1 and R1) of PCUs were used to maximize S/N in the 3-20 keV energy band.  For the observations after May 2000, for which the propane layer for PCU0 was lost, data obtained from PCU0 was not used in the spectral analysis.

Out of 167 total pointings, we excluded X-ray eclipses (due to low count statistics) and used data obtained from $\sim 130$ pointings with a total exposure of $\sim 250$ ksec for the timing analysis. Other than the public observations, the principal investigators of these observations are S. Eikenberry, W. Heindl, D. Chakrabarty, G.W. Clark and J. Deeter. During the observations between 1998 October 16 and 1998 November 24, with an exposure of $\sim$ 23 ks, the data were contaminated due to the outburst of the nearby transient X-ray pulsar XTEJ0111.2-7317 (Chakrabarty et al 1998). We did not use these observations in the X-ray spectral analysis. In order to obtain the timing solution of the source, we used observations with the proposal number 10139 covering a time span of about 6 days. These observations are more suitable than other observations to obtain the timing solution, since they are the only set of observations that continously cover near two orbital periods.

\section{Timing Analysis} 

For the timing analysis, we generated lightcurves (3-20 keV) with 0.035s timing resolution using Good Xenon data. These lightcurves were background corrected with a 16s binned background lightcurve constructed using background estimator models from the RXTE team (URL: ftp://legacy.gsfc.nasa.gov/xte /calib\_data/pca\_bkgd/). The resulting lightcurves were corrected to the barycenter of the Solar system. A template pulse profile from each RXTE observation was constructed by folding the data on the period which had the greatest power in the periodogram. Pulse arrival times were found by cross-correlating the pulse profiles obtained from $\sim 200$s long segments with the template pulse profile. Both the template and cross-correlated pulse profiles consisted of 20 phase bins. Crosscorrelation was performed by using the harmonic representation of pulse profiles (Deeter $\&$ Boynton 1985). In order to obtain pulse arrival times, the pulse profiles were expressed in terms of harmonic series and cross-correlated with the template pulse profile. 

\subsection{Timing Solution}

Initially, we found pulse arrival times obtained from a series of observations with a time span of $\sim 6$ days around MJD 50324.7. These pulse arrival times can be fitted to an expression to obtain timing solution (Deeter, Boynton, $\&$ Pravdo 1981). We firstly assumed a circular orbit (e=0) and therefore this expression became,

\begin{equation}
\delta t = \frac{\delta P}{P} (t-t_{0})
+\frac{1}{2}\frac{\dot P}{P}(t-t_{0})^{2}+x sin(l).
\end{equation}

Here t$_{0}$ is the mid-time of the observation;
 $\delta P$ is the deviation from mean pulse period;
 $\dot P$ is the time derivative of the pulse period;
$x=a/c \sin(i)$ is the light traveltime
 for projected semimajor axis
(where i is the inclination angle between the line of sight
and the orbital angular momentum vector);
$l=2\pi (t-T_{\pi/2})/P_{orbit}+\pi/2$ is the mean orbital longitude
at t; $T_{\pi/2}$ is the epoch when the mean orbital longitude is equal to
90 $^{\circ}$; $P_{orbit}$ is the orbital period of the system. In order to improve the fit, we used the whole expression containing terms corresponding to an eccentric orbit as  

\begin{equation}
\delta t = \frac{\delta P}{P} (t-t_{0})
+\frac{1}{2}\frac{\dot P}{P}(t-t_{0})^{2}+x sin(l)
-\frac{3}{2}xesin(w)+\frac{1}{2}xecos(w)sin(2l)
-\frac{1}{2}xesin(w)cos(2l).
\end{equation}

In Equation 2, e is the eccentricity and w is the
 longitude of periastron. The periodic trend of the pulse arrival times yields
an eccentric orbit (e=0.00089(6)) with an orbital period of
3.89220909(4) days (numbers in paranthesis denote $1\sigma$ uncertainties in the least significant figures hereafter). Table 1 presents the timing solution of
SMC X-1. Figure 1 presents the pulse arrival times after the
 removal of the quadratic trend (or intrinsic -$\dot P$) together with
the residuals of circular (e=0) and eccentric orbit models respectively. 

In Table  2, we present the orbital epoch ($T_{\pi/2}$) measurements from
different observatories and orbital cycle number (N).
In Figure 2, we present observed minus calculated values
of orbital epochs
($T_{\pi/2}-n<P_{orbit}>-<T_{\pi/2}-n<P_{orbit}>>$) relative to the
constant orbital period ($<P_{orbit}>= 3.892188$ days). It should be noted that the rightmost point indicated as "RXTE Point" is our result and other points were already published before (see Figure 7 in Wojdowski et al. 1998).
A quadratic fit to the epochs from all experiments yielded
an estimate of the rate of period change
$\dot P_{orb}/P_{orb} =-3.402(7) \times 10^{-6}$ yr$^{-1}$.

\begin{table}
\caption{Timing Solution of SMC X-1}
\centering
\begin{tabular}{l|l}\hline\hline
parameters                   & Model (This Work)  \\ \hline
Timing Epoch (MJD)  & 50326.62356961(9)  \\
$\nu$ (Hz)          & 1.413630801(4) \\
$\dot{\nu}$ ($10^{-11}$ Hz.s$^-1$)    &  3.279(5) \\
Orbital Period (days)        & 3.89220909(4) \\
a/c sin I (lt-s)             & 53.4876(9)  \\
Orbital Epoch                & MJD 50324.691861(8)    \\
$\dot{P_{orbit}}/P_{orbit}$ ($10^{-6}$ yr$^{-1}$) & -3.402(7)      \\
Eccentricity                 & 0.00089(6)         	\\
w (longitude of periastron)  & 166(12)          \\    \hline 
\end{tabular}
\label{Table1 -- Parameters}
\end{table}

\begin{table}
\caption{Orbital Epoch Measurements of SMC X-1}
\begin{center}
\begin{tabular}{l|l|l|l}\hline\hline 
$T_{\pi/2}$ (MJD) & N & Observatory & Reference \\ \hline
40963.99(2)     & -481   & Uhuru   & 1,2 \\
42275.65(4)     & -144   & Copernicus & 1,3 \\
42836.1828(2)   &  0     & SAS 3   & 1,4 \\
42999.6567(16)   & 42    & Ariel V & 1,5 \\
43116.4448(22)   & 72    & Cos B   & 1,6 \\
46942.47237(15)  & 1055  & Ginga   & 1,7 \\
47401.744476(7) &  1173  & Ginga   & 1,7 \\
47740.35906(3)  &  1260  & Ginga   & 1,7 \\
48534.34786(35)  & 1464  & ROSAT   & 1 \\
49102.59109(82)  & 1610  & ASCA	   & 1 \\
49137.61911(50)  & 1619  & ROSAT   & 1 \\
50091.170(63)    & 1864  & RXTE	   & 1 \\
50324.691861(8) &  1924  & RXTE	   & 8  \\ \hline			   
\end{tabular}
\end{center}
{\small{References: (1) Wojdowski et al. 1998; (2) Schreier et al. 1972; (3) Tuohy \& Rapley 1975; (4) Primini et al. 1977; (5) Davison 1977; (6) Bonnet-Bidaud et al. 1981; (7) Levine et al. 1993; (8) This work}}
\end{table}

\begin{figure}
\begin{center}
\psfig{file=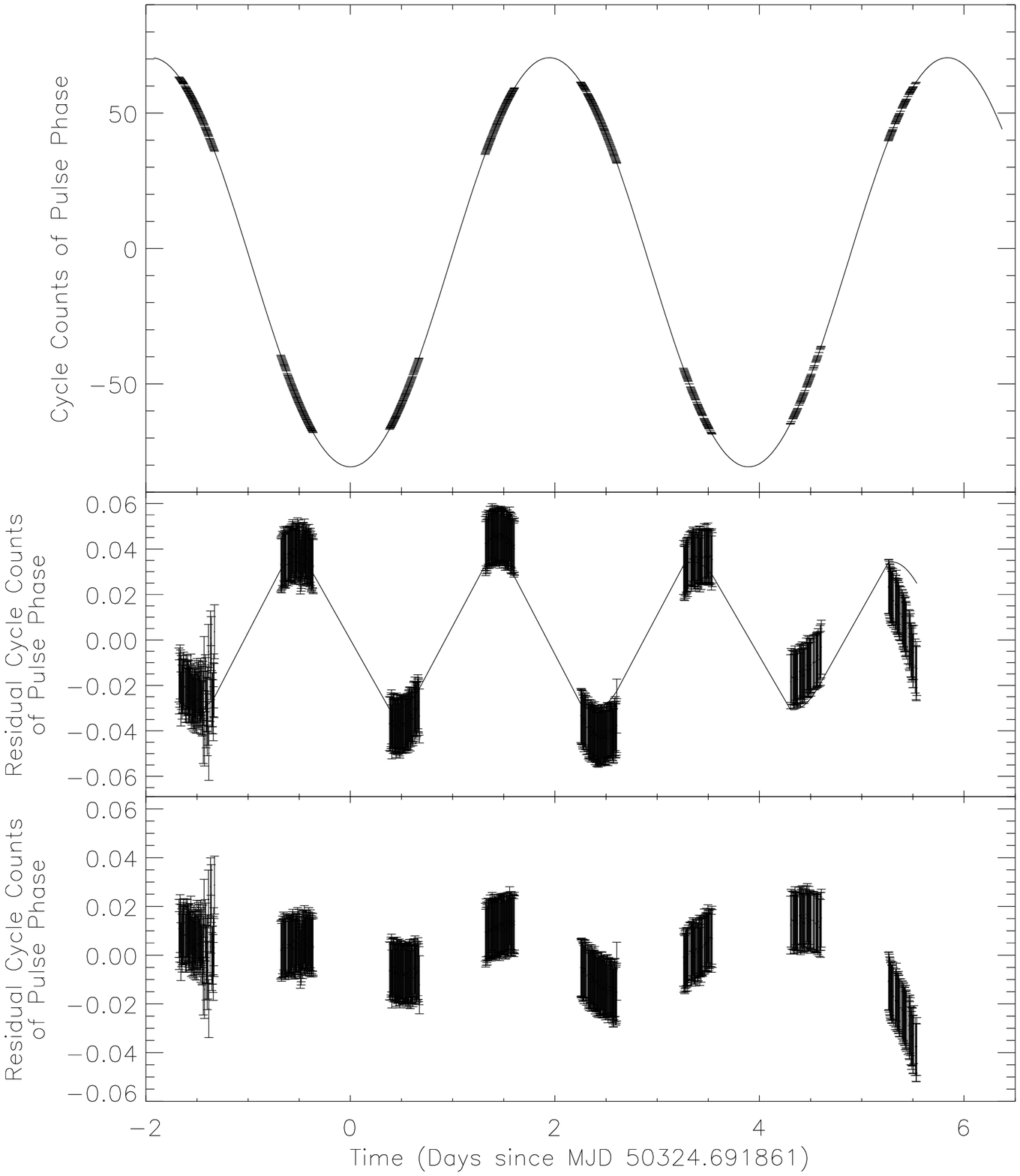,height=8cm,width=13cm,angle=0}
\end{center}
\small{Figure 1 -- Arrival times (pulsation cycles) (top panel), and residuals after fitting arrival times to a circular (middle panel) and elliptical orbital model (lower panel) obtained from the observations with a time span of about 6 days around 30 August 1996.}
\end{figure}

\begin{figure}
\begin{center}
\psfig{file=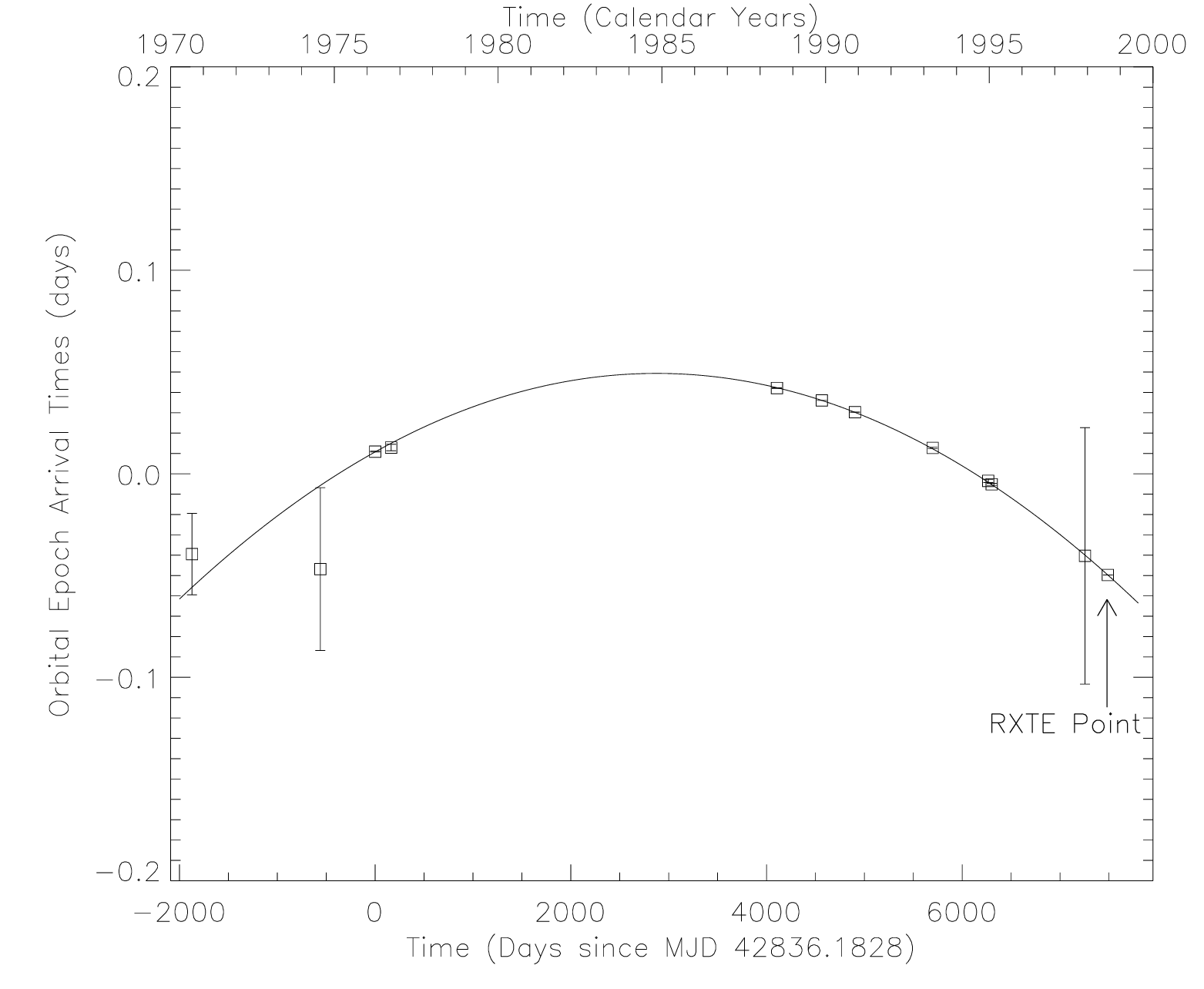,height=8cm,width=13cm,angle=0}
\end{center}
\small{Figure 2 -- Orbital Epoch Arrival Times of SMC X-1 with a time span of about 30 years. From the quadratic fit to the data, an orbital period decay with $\dot{P_{orbit}}/P_{orbit}=-3.402(7)\times10^{-6}$ yr$^{-1}$ is found.}
\end{figure}

\begin{figure}
\begin{center}
\psfig{file=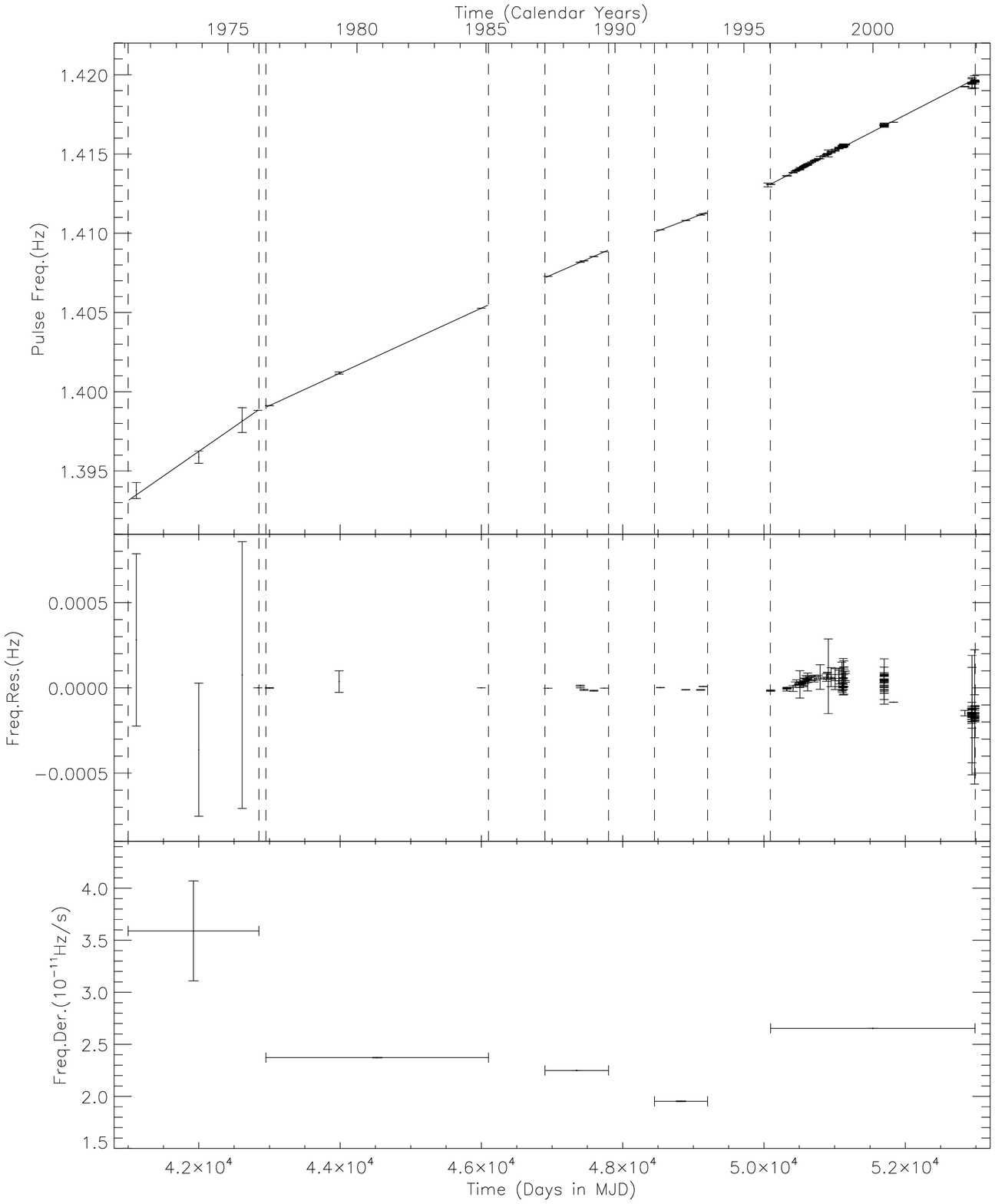,height=12cm,width=13cm,angle=0}
\end{center}
\small{Figure 3 -- {\bf{(top)}} and {\bf{(middle)}} Pulse frequency history and residuals of SMC X-1 after a linear fit obtained from 5 different intervals listed in Table 4. {\bf{(bottom)}} Frequency derivative history. Dashed lines in top and middle panels and horizontal error bars in the bottom panel indicate time intervals in which linear fits were performed. The last (rightmost) interval almost consists of pulse frequency values obtained from our analysis.}
\end{figure}

\begin{table}[ht]
\caption{Pulse Frequency History of SMC X-1} 
\begin{center} 
\begin{tabular}{c c c c} 
\hline\hline
Proposal ID or Observatory & MJD (days) & Pulse Period$^1$ & References\\\hline 
Uhuru         & 41114.200   & 1.39377(51)     & Henry $\&$ Schreier, 1977 \\  
Aerobee       & 41999.600   & 1.3959(4)       & Yentis et al 1977 \\  
Apollo-Soyuz  & 42613.799   & 1.3982(8)	      & Henry $\&$ Schreier, 1977  \\  
SAS 3         & 42836.183   & 1.3988247(1)   & Wojdowski et al 1998 \\  	    
Ariel V       & 42999.6567  & 1.3991225(23)   & Wojdowski et al 1998 \\    
Einstein      & 43985.907   & 1.401180(63)    & Wojdowski et al 1998 \\	   
EXOSAT        & 45998.500   & 1.4052680772(8)   & Kunz et al 1993 \\ 		
Ginga         & 46942.4724  & 1.407277083(71)   & Levine et al 1993 \\		
HEXE          & 47399.500   & 1.4081806(29)     & Kunz et al 1993  \\           
Ginga         & 47401.744   & 1.408171707(30)   & Levine et al 1993  \\	      
HEXE          & 47451.500   & 1.4082567(30)     & Kunz et al 1993  \\           
HEXE          & 47591.000   & 1.4085229(30)     & Kunz et al 1993  \\	      
Ginga         & 47740.3591  & 1.408827913(64)   & Levine et al 1993 \\	       
ROSAT 1       & 48534.3479  & 1.41021063(68)    & Wojdowski et al 1998 \\	
ROSAT 2       & 48892.4191  & 1.4108022(12)      & Wojdowski et al 1998  \\	
ASCA          & 49102.5911  & 1.4111556(18)      & Wojdowski et al 1998 \\	
ROSAT 3       & 49137.6191  & 1.4112350(10)      & Wojdowski et al 1998 \\	
ROSAT         & 50054.000   & 1.41305(12)        & Kahabka $\&$ Li 1999 \\      
RXTE          & 50091.170   & 1.41308143(16)     & Wojdowski et al 1998 \\	00011 & 50093.048   & 1.4130802(13)  &  This paper  \\		      
00011 & 50093.115   & 1.4130802(39) &  This paper \\  		      
00011 & 50093.181   & 1.4130786(14)  &  This paper \\		      
10139 & 50323.016   & 1.41361728(15) &  This paper \\ 		   	      
10139 & 50323.349   & 1.4136112(52) &  This paper \\  		       
10139 & 50324.010   & 1.413626129(76)  &  This paper \\		      
10139 & 50325.069   & 1.41362903(34)  &  This paper \\		      
10139 & 50326.009   & 1.413627028(64)    &  This paper \\  		      
10139 & 50327.268   & 1.4136316(18)       &  This paper \\  		      
10139 & 50327.941   & 1.413637100(64)    &  This paper \\ 
10139 & 50328.993   & 1.413640077(64)    &  This paper \\		       
10139 & 50329.940   & 1.413637060(64)    &  This paper \\ 		  
20146-20109-20417 & 50411.87    & 1.413823(17)    & This paper  \\ 	  
20146-20109-20417 & 50442.90    & 1.413918(16)    & This paper  \\ 	      
BeppoSAX-1        & 50460.900   & 1.413971(16)    & Naik $\&$ Paul et al 04	\\ 	      
\hline 
\end{tabular}
\end{center}
{\small{$^1$ {errors indicated in the paranthesis gives the value for each pulse period in 1$\sigma$}}} \\
\end{table}

\begin{table}[ht]
\begin{center} 
\begin{tabular}{c c c c} 
\hline\hline
Proposal ID or Observatory & MJD (days) & Pulse Period$^1$  & References\\\hline 				  
20146-20109-20417 & 50489.366   & 1.4140261(14)     & This paper	\\     	
20146-20109-20417 & 50504.524   & 1.4140567(16)    & This paper\\ 	  
20146-20109-20417 & 50505.169   & 1.4140587(18)     & This paper\\ 		 
BeppoSAX-2        & 50507.600   & 1.414067(8)    & Naik $\&$ Paul et al 04 \\ 	 
20146-20109-20417 & 50531.806   & 1.4141297(10)    & This paper	\\ 		 
20146-20109-20417 & 50535.679   & 1.4141399(14)     & This paper\\ 		 
20146-20109-20417 & 50551.398   & 1.4141789(10)     & This paper\\ 		 
20146-20109-20417 & 50556.346   & 1.414180(15)    & This paper\\ 		 
BeppoSAX-3        & 50562.100   & 1.414199(14)   & Naik $\&$ Paul, 2004\\     
20146-20109-20417 & 50566.861   & 1.4142307(9)    & This paper\\ 		 
20146-20109-20417 & 50579.511   & 1.414251(16)    & This paper\\ 		 
20146-20109-20417 & 50583.009   & 1.4142586(5)   & This paper\\        		 
20146-20109-20417 & 50583.030   & 1.4142597(14)     & This paper\\ 		 
ROSAT             & 50593.799   & 1.41430(2)     & Kahabka $\&$ Li, 1999 \\ 	 
20146-20109-20417 & 50597.960   & 1.414306(2)  & This paper\\ 		  
20146-20109-20417 & 50598.980	& 1.4143059(14)     & This paper\\ 		 
20146-20109-20417 & 50609.319	& 1.414340(5)   & This paper\\	  
20146-20109-20417 & 50617.757	& 1.4143399(96)     & This paper\\   
20146-20109-20417 & 50617.898	& 1.41433(5)    & This paper\\	  
20146-20109-20417 & 50646.787	& 1.414420(7)     & This paper\\	  
20146-20109-20417 & 50675.866	& 1.414488(13)     & This paper\\	  
20146-20109-20417 & 50711.833	& 1.41457(2)      & This paper\\	  
20146-20109-20417 & 50738.053	& 1.414631(10)     & This paper\\   
20146-20109-20417 & 50767.442	& 1.414706(10)     & This paper\\	  
20146-20109-20417 & 50795.116	& 1.41477(7)      & This paper\\                
30125 & 50850.541   & 1.414891(5)   & This paper\\     			      
30125 & 50883.272   & 1.41497(2)  & This paper \\ 	       
ROSAT & 50898.200   & 1.41501(2)	& Kahabka $\&$ Li, 1999 \\       
30125 & 50910.965   & 1.4150(2) & This paper\\ 	      
30125 & 50949.407   & 1.415116(5)   & This paper\\ 		      
30125 & 50949.751   & 1.41512(6)  & This paper\\ 		      
30125 & 50950.048   & 1.415128(11)   & This paper\\ 		      
30125 & 51007.374   & 1.41524(6)  & This paper\\ 		      
30125 & 51007.975   & 1.415242(7)   & This paper\\ 		      
30125 & 51060.725   & 1.41538(5)  & This paper\\ 		      
30125 & 51060.791   & 1.41538(4)  & This paper\\ 	      
30125 & 51061.265   & 1.415376(18)   & This paper\\ 
\hline 
\end{tabular}
\end{center}
{\small{$^1$ {errors indicated in the paranthesis gives the value for each pulse period in 1$\sigma$}}} \\
\end{table}

\begin{table}[ht]
\begin{center}
\begin{tabular}{c c c c} 
\hline\hline
Proposal ID or Observatory & MJD (days) & Pulse Period$^1$  & References\\\hline
30090-30125 & 51102.221   & 1.41548(9) & This paper \\			    
30090-30125 & 51106.083   & 1.41548(2) & This paper \\ 			    
30090-30125 & 51108.060   & 1.415432(18)  & This paper \\		     
30090-30125 & 51109.781   & 1.41543(3) & This paper \\			     
30090-30125 & 51111.866   & 1.41544(2) & This paper \\ 			    
30090-30125 & 51113.711   & 1.415443(4)  & This paper \\ 		    
30090-30125 & 51115.709   & 1.415446(6)  & This paper \\ 		    
30090-30125 & 51117.607   & 1.415451(20)  & This paper \\ 		    
30090-30125 & 51119.586   & 1.415455(4)  & This paper \\		     
30090-30125 & 51121.160   & 1.415509(2)  & This paper \\ 		    
30090-30125 & 51121.515   & 1.4155127(14)   & This paper \\ 		    
30090-30125 & 51121.637   & 1.41552(11) & This paper \\			     
30090-30125 & 51121.724   & 1.41551(4) & This paper \\ 	    
30090-30125 & 51125.443   & 1.41552(7) & This paper \\ 		    
30090-30125 & 51127.234   & 1.415528(17)  & This paper \\ 		    
30090-30125 & 51129.376   & 1.41553(10) & This paper \\ 		    
30090-30125 & 51131.300   & 1.41553(5) & This paper \\ 		    
30090-30125 & 51133.301   & 1.41555(6) & This paper \\ 			    
30090-30125 & 51151.014   & 1.41559(3) & This paper \\ 			    
30090-30125 & 51151.416   & 1.41558(3) & This paper \\ 		     
40064 & 51699.405  & 1.416816(9)   & This paper \\ 		     
40064 & 51699.666  & 1.416824(9)   & This paper \\ 		     
40064 & 51699.736  & 1.41681(3)  & This paper \\      
40064 & 51699.876  & 1.416812(16)   & This paper \\ 		     
40064 & 51700.172  & 1.41681(6)  & This paper \\ 		     
40064 & 51700.233  & 1.416809(22)   & This paper \\		     
40064 & 51700.383  & 1.416809(20)  & This paper \\ 		     
40064 & 51700.593  & 1.41681(9)   & This paper \\ 		     
40064 & 51700.663  & 1.416808(47)   & This paper \\		      
40064 & 51700.803  & 1.416810(14)   & This paper \\ 		     
40064 & 51700.896  & 1.416807(57)  & This paper \\ 		     
40064 & 51701.152  & 1.416814(45)  & This paper \\ 		     
40064 & 51701.315  & 1.416817(3)   & This paper \\ 		     
40064 & 51701.590  & 1.41682(13)  & This paper \\ 		     
40064 & 51701.660  & 1.416819(43)  & This paper \\		      
40064 & 51701.869  & 1.416822(33)  & This paper \\		     
40064 & 51701.957  & 1.4168218(3)     & This paper \\ 
\hline 
\end{tabular}
\end{center}
{\small{$^1$ {errors indicated in the paranthesis gives the value for each pulse period in 1$\sigma$}}} \\
\end{table}

\begin{table}[ht]
\begin{center} 
\begin{tabular}{c c c c} 
\hline\hline
Proposal ID or Observatory & MJD (days) & Pulse Period$^1$ & References\\\hline 
Chandra   & 51833.808 & 1.4170032(6)  & Vrtilek et al 2005\\ 			
INTEGRAL  & 52843     & 1.419253(16)   & McBride et al 2007 \\		      
80078 & 52939.611  & 1.419470(5)  & This paper \\ 		     
80078 & 52939.680  & 1.419472(6)  & This paper \\ 	     		     
80078 & 52939.749  & 1.4194708(19)   & This paper \\ 		     
80078 & 52940.527  & 1.4194728(50)  & This paper \\ 		     
80078 & 52940.596  & 1.419473(34) & This paper \\ 		     
80078 & 52940.665  & 1.41947(3) & This paper \\ 		     
80078 & 52940.734  & 1.4194722(15)   & This paper \\ 		     
80078 & 52940.805  & 1.419471(3)  & This paper \\ 		     
80078 & 52940.958  & 1.419470(13)  & This paper \\ 		     
80078 & 52941.304  & 1.419468(50) & This paper \\ 		        
80078 & 52941.373  & 1.419465(38) & This paper \\ 		     
80078 & 52941.442  & 1.419464(36) & This paper \\ 		     
80078 & 52941.511  & 1.419464(11)  & This paper \\ 		     
80078 & 52941.813  & 1.419460(4)  & This paper \\ 		     
80078 & 52942.289  & 1.419462(2)  & This paper \\ 		     
80078 & 52942.565  & 1.419466(33)  & This paper \\		     
80078 & 52942.652  & 1.419469(76)  & This paper \\ 		     
80078 & 52942.825  & 1.41947(28) & This paper \\ 		     
80078 & 52942.891  & 1.41947(35) & This paper \\ 		     
80078 & 52943.688  & 1.419481(6)   & This paper \\		      
80078 & 52981.226  & 1.4195(4) & This paper \\ 		     
80078 & 52981.527  & 1.41955(13)  & This paper \\		      
80078 & 52982.064  & 1.419578(10)   & This paper \\ 		     
80078 & 52983.496  & 1.41958(3)  & This paper \\ 		     
80078 & 52984.033  & 1.419572(6)   & This paper \\		      
80078 & 52984.360  & 1.419570(45)  & This paper \\		      
80078 & 52984.480  & 1.419567(36)  & This paper \\		      
80078 & 52984.687  & 1.419566(12)   & This paper \\		      
80078 & 52984.890  & 1.419566(16)   & This paper \\		      
80078 & 52985.326  & 1.419571(7)   & This paper \\ 		     
80078 & 52986.311  & 1.419586(25)  & This paper \\		     
80078 & 52987.433  & 1.419587(39)  & This paper \\		     
80078 & 52987.640  & 1.419585(43)  & This paper \\ [1ex] 
\hline 
\end{tabular}
\end{center}
{\small{$^1$ {errors indicated in the paranthesis gives the value for each pulse period in 1$\sigma$}}} \\
\end{table}

\subsection{Pulse Period History}

For the other observations between MJD 50093 and MJD 52988 (each having $\sim 1-2$ksec exposure), we roughly obtained pulse periods using the periodogram and then refined these periods by the use of a linear fit to the arrival times. Errors were estimated using the scattering of arrival time values (i.e. errors were related to the errors of the linear coefficient of the fit). These periods were corrected for the binary motion of the pulsar using the orbital parameters listed in Table 1. We present the whole pulse period history of the source in Table 3 and Figure 3 including our results. From Figure 3, it is evident that the source spins up continuosly between MJD $\sim40000$ and MJD $\sim53000$ with a varying spin-up rate. In Table 4, average spin-up rate values of the source are listed. 

\begin{table}
\caption{Long Term Spin-up Rate Values of SMC X-1}
\begin{center}
\begin{tabular}{l|l}\hline\hline 
Interval (MJD-MJD) & Spin-up Rate ($10^{-11}$ Hz s${-1}$) \\ \hline
41000-42850 & 3.59(48) \\
42950-46100 & 2.3718(15) \\
46900-47800 & 2.2491438(1) \\
48450-49200 & 1.9528(38) \\
50093-52988 & 2.65343816(7) \\ \hline
\end{tabular}
\end{center}
\end{table}

\section{Spectral Analysis}
 
The same 130 RXTE PCA observations used for the timing study were also used for the spectral analysis except those between 1998 October 16 and 1998 November 24 which were contaminated due to the outburst of the nearby transient X-ray pulsar XTEJ0111.2-7317 (Chakrabarty et al 1998). We used the Standard-2 mode data, which provides 128 channel spectra at 16 sec time resolution. Spectrum, background and response matrix files were created using FTOOLS 6.3 data analysis software. We used background subtracted spectra in our analysis. Energy channels corresponding to the 3-25 keV energy range were used to fit the spectra. We ignored photon energies lower than 3 keV and higher than 25 keV and 1$\%$ systematic error was added to the errors (see Wilms et al. 1999; Coburn et al. 2000).

To fit the spectra, we used a power law model with an high energy cutoff and a Gaussian component peaking at 6.7 keV (Angelini et al 1991). We also tried to add a partial covering absorption component (Neilsen et al 2004), but adding this model component did not improve the fit.
   
\begin{table}[ht]
\caption{Sample Spectral Parameters from Two Different Observations} 
\centering 
\begin{tabular}{c c c} 
\hline\hline 
Obs ID & 10139-01-01-00 & 20146-06-02-00 \\  
\hline 
MJD & 50323.349  & 50442.905 \\
nH$(10^{22}cm^{-2})$ & 12.039$\pm$0.748 & 2.214$\pm$0.290\\
Powerlaw Index & 1.184$\pm$0.043 & 1.176$\pm$0.074 \\  
$E_{cutoff}$(keV) & 13.697$\pm$3.424 & 6.595$\pm$1.649 \\   
$E_{fold}$ (keV) & 11.765$\pm$4.118 & 10.020$\pm$3.507 \\ 
$({\chi}^2_{\nu})$ (degrees of freedom) & 0.868(47) & 1.055(47) \\
Obserbed flux $(10^{-10}$ ergs cm$^{-2}$ s$^{-1})$(3-25 keV) & $2.3866^{+0.0282}_{-0.102}$ & $20.673^{+0.043}_{-0.157}$ \\ 
Unabsorbed flux $(10^{-10}$ ergs cm$^{-2}$ s$^{-1})$(3-25 keV) & $2.8281^{+0.002}_{-0.012}$ & $21.553^{+0.039}_{-0.078}$ \\ [1ex]
\hline 
\end{tabular}
\label{Table 4 -- example two data} 
\end{table}

Table 5 shows best fit parameters of the spectral model for two sample observations. In general, we found that power law index, high energy cut-off and e-fold energy does not show variations with time, orbital phase and X-ray flux within the errors. We found that Hydrogen column density gets higher when X-ray flux is lower (see Figure 4).

\begin{figure}
\begin{center}
\psfig{file=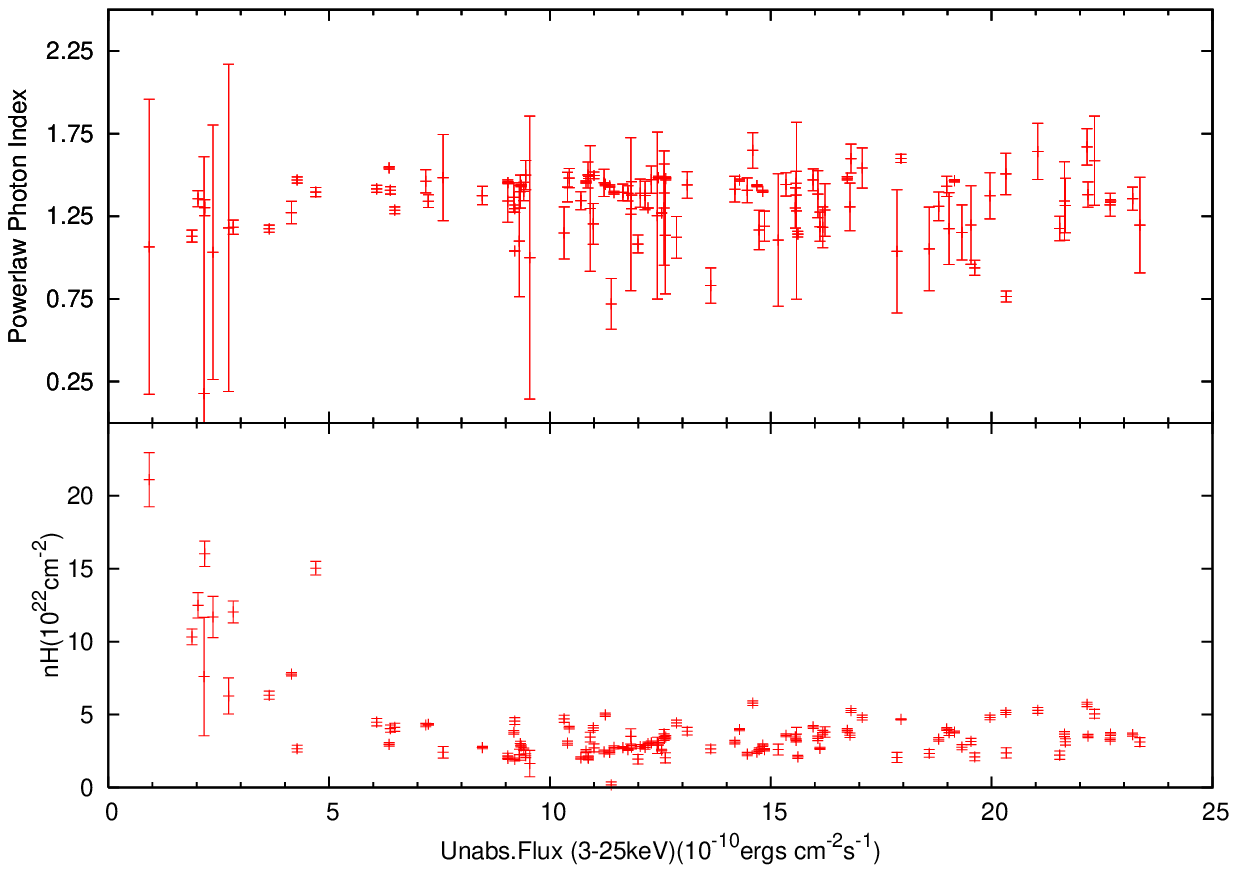,height=8cm,width=13cm,angle=0}
\end{center}
\small{Figure 4 -- Dependence of photon index and $n_H$ on 3-25 keV unabsorbed X-ray flux.}
\end{figure}

\section{Discussion}

In this paper, we presented timing and spectral analysis of RXTE-PCA observations of SMC X-1 with a time span of about 8 years. From timing analysis, we revised timing solution of the source and resolved an eccentricity value. We also confirmed the orbital decay reported before. Our timing analysis helped us to construct a $\sim 30$ year long pulse period history of the source (see Figure 3).  From the spectral analysis, we found that all spectral parameters except Hydrogen column density showed no significant variation with time and X-ray flux. 

Figure 1 demonstrates that the eccentric orbit model is a better fit compared to the circular orbital model. The timing solution of the source revealed that the binary orbit has an eccentricity of 0.00089(6). Wojdowski et al. (1998) had found a circular orbit solution and Levine et al. (1993) had only found an upper limit of 0.00004 for eccentricity. Our present value is more than 20 times greater than the upper limit found by Levine et al. 1993. This significant eccentricity value should be verified using future monitoring observations. 

From timing analysis, we obtained a new orbital epoch of SMC X-1 from RXTE observations. Using this new epoch and previous results (Wojdowski et al. 1998), we found that there is an orbital decay with $\dot{P_{orbit}}/P_{orbit}$ of $-3.402(7)\times10^{-6}$ yr$^{-1}$. This value is similar to the values found by Wojdowski et al. (1998) and Levine et al. (1993). Levine et al. (1993) proposed that the major cause of change in the orbital period is tidal interactions as for the case in Cen X-3 (Kelley et al. 1983), and LMC X-4 (Levine et al. 2000). SMC X-1 is unlike Her X-1 for which the mass accretion were thought to be primarily related to the orbital period decay (Deeter et al. 1991). 

From Figure 3, it is seen that SMC X-1 spins up continuosly for $\sim 30$ years. We found that average long term spin-up rate of the source between MJD 50093.048 and MJD 52987.640 (whole time span of our analysis) is $2.65343816(7)\times 10^{-11}$ Hz s$^{-1}$. 

Spin-up rate of a source with a persistent accretion disk can be expressed as

\begin{equation}
\dot{\nu}\simeq 2.2\times10^{-12}\mu_{30}^{2/7}m_x^{-3/7}R_6^{6/7}I_{45}^{-1}L_{37}^{6/7} {\rm{Hz\,s}}^{-1},
\end{equation}

\noindent{where $\dot{\nu}$ is first time derivative of the spin frequency, $\mu_{30}$ is the magnetic moment of the neutron star in units of $10^{30}$ Gauss cm$^{3}$, $m_x$ is the mass of the neutron star in units of solar mass, $R_6$ is the radius of the neutron star in units of cm, $I_{45}$ is the moment of inertia of the neutron star in terms of $10^{45}$ g cm$^2$, and $L_{37}$ is the luminosity of the neutron star in  units of $10^{37}$ erg s$^{-1}$ (Ghosh\& Lamb, 1979). Using a distance value of 61kpc (Keller\& Wood 2006) and a flux of $1.2\times 10^{-9}$ ergs cm$^{-2}$ s$^{-1}$ (see Figure 4), luminosity of SMC X-1 is $\sim 5.5\times 10^{38}$ erg s$^{-1}$. Assuming magnetic moment of $10^{30}$ Gauss cm$^3$, mass of $1.4M_{\odot}$, radius of $10^{6}$ cm, moment of inertia of $10^{45}$ g cm$^{2}$, we find $\dot{\nu}$ to be $\sim 5.9\times 10^{-11}$ Hz s$^{-1}$. This value is of the order of the observed spin-up rate value of the source.}  
 
From Figure 3 and Table 4, it is evident that the long term spin-up rate of SMC X-1 varies within a factor of 2. From Equation 3, this variation may -in principle- be related to a change in X-ray luminosity of the source. However, previous X-ray observations of the source were performed by different X-ray observatories and/or in different energy bands. These observations are sparsely distributed and average X-ray flux of these observations are not likely to represent an average X-ray flux for the intervals for which the pulse frequency derivatives were found. Moreover, X-ray flux variations are also affected by the super-orbital cycle, period of which is not as stable as that of Her X-1 (Clarkson et al. 2003). Therefore, it is not possible to test whether variation in long term spin-up rate is related to a change in X-ray luminosity or not.

From Table 1, spin-up rate obtained from $\sim 6$days long observation around MJD 50324.7 is $3.279(5)\times 10^{-11}$ Hz s$^{-1}$. This is about 20$\%$ greater than the long term spin-up rate between MJD 50093 and 52988. Assuming that the pulse frequency variations can be explained in terms of random walk model in pulse frequency (Baykal \& Ogelman 1993), one can express the pulse frequency derivative variations as $<\Delta\dot{\Omega}^{1/2}>=\sqrt{{S}\over{t}}$ where S is the noise strength and t is the time interval at which the pulse frequency derivatives are observed. Therefore it is natural to observe high pulse frequency derivative fluctuations at shorter time scales. It is important to note that we refined pulse period values fitting arrival times each obtained from $\sim 200$s long intervals, therefore we had to have at least a few arrival times to obtain spin period accurately. So, our shortest timescale to obtain pulse periods is of the order of a single RXTE observation which is $\sim 1-2$ksec long.   

From the spectral analysis, we found that all of the spectral parameters except Hydrogen column density did not vary significantly. Hydrogen column density was found to be higher as X-ray flux gets lower.  Increase in Hydrogen column density with a decrease in X-ray flux is also observed in Her X-1 (Inam\& Baykal, 2005) and may be due the fact that soft absorption becomes stronger whenever there is a partial obscuration of the neutron star due to the X-ray eclipses and warping of the accretion disk. The increase in Hydrogen column density may also be an artifact of the simple absorbed power law model. Paul et al. (2002) showed that SMC X-1 has soft excess especially for energies $\aplt 3$ keV. Although we used photon energies greater than 3keV in our analysis, tail of a soft spectral compenent may affect low energies in our analysis and as a result we might have misidentified corresponding changes in energy spectrum as  variations in Hydrogen column density parameter.  

\section*{Acknowledgements}
We acknowledge support from EU FP6 Transfer
of Knowledge Project "Astrophysics of Neutron Stars"
(MTKD-CT-2006-042722).

\section*{References}
Angelini L., Stella L. and White N.E., 1991, ApJ, 371, 332 \\

Baykal A., Ogelman H., 1993, A\&A, 267, 119 \\

Bonnet-Bidaud J. M., van der Klis M., 1981, A\&A, 97, 134 \\

Chakrabarty D., Levine A. M., Clark G. W., Takeshima T., 1998, IAUC, 7048 \\

Clarkson W.I., Charles P.A., Coe M.J., Laycock S., Tout M.D., Wilson C.A., 2003, MNRAS, 339, 447 \\

Coburn W. et al. 2000, ApJ, 543, 351 \\

Davison P. J. N., 1977, MNRAS, 179, 15 \\

Deeter J. E., Boynton P. E., Pravdo S. H., 1981, ApJ, 247, 1003 \\

Deeter J. E., Boynton P. E., 1985, in Proc. Inuyama Workshop on Timing
Studies of X-Ray Sources, ed. S. Hayakawa \& F. Nagase (Nagoya: Nagoya
Univ.), 29 \\

Deeter J.E., Boynton P.E., Miyamoto S. et al. 1991, ApJ, 383, 324 \\

Galache J. L., Corbet R. H. D., Coe M. J., Laycock S., Schurch M. P. E., Markwardt C., Marshall F. E., Lochner J., 2008, ApJS, 177, 189 \\

Ghosh P., Lamb F.K., 1979, ApJ, 234, 296 \\

Gruber D. E., Rothschild R. E., 1984, ApJ, 283, 546 \\

Inam S. C., Baykal A., 2005, MNRAS, 361, 1393 \\

Henry P., Schrier E., 1977, ApJ, 212, L13 \\  

Jahoda K., Swank J. H., Giles A. B., Stark M. J., Strohmayer T., Zhang W., Morgan, E. H., 1996, SPIE, 2808, 59 \\

Kahabka P., Li X.D., 1991, A$\&$A, 345, 117 \\ 

Keller S.C., Wood P.R., 2006, ApJ, 642, 834 \\

Kelley R.L., Rappaport G., Clark G.W., Petro L.D. 1983, ApJ, 268, 790 \\

Kunz M., Gruber D.E., Kendziorra E. et al. 1993, A$\&$A, 268, 116  \\  
	       
Levine A., Rappaport S., Deeter J.E., Boynton P.E., Nagase F., 1993, ApJ, 410, 328 \\

Levine A.M., Rappaport S.A., Zojcheski G., 2000, ApJ, 541, 194

Lucke R., Yentis D., Friedman H., Fritz G., Shulman S., 1976, ApJL, 206, 25 \\ 

McBridge V.A. ,Coe M.J., Bird A.J., Dean A.J., Hill A.B., McGowan K.E., Schurch M.P.E., Udalski A., Soszynski I., Finger M., Wilson C.A.,
Corbet R.H.D., Negueruela I., 2007, MNRAS, 382, 743 \\ 

Naik S. and Paul B., 2004, A$\&$A, 418, 655 \\ 	

Neilsen J., Hickox R. C., Vrtilek S. D., 2004, ApJL, 616, 135 \\

Paul B., Nagase F., Endo T., Dotani T., Yokogawa J., Nishiuchi M., 2002, ApJ, 579, 411 \\

Price R. E., Groves D. J., Rodrigues R. M., Seward F. D., Swift C. D., Toor A., 1971, ApJL, 168, 7 \\

Primini F.A., 1977, PhD thesis submitted to Massachusetts Institute of Technology \\

Reynolds A. P., Hilditch R. W., Bell S. A., Hill G., 1993, MNRAS, 261, 337 \\

Schreier E., Giacconi R., Gursky H., Kellogg E., Tananbaum H., 1972, ApJL, 178, 71 \\

Trowbridge, S., Nowak M. A., Wilms, J., 2007, ApJ, 670, 624 \\

Tuohy I. R., Rapley, C. G., 1975, ApJL, 198, 69 \\	

van der Meer A., Kaper L., van Kerkwijk M. H., Heemskerk M. H. M., van den Heuvel E. P. J., 2007, A\&A, 473, 523 \\

Vrtilek S.D., Raymond J.C., Boroson B., McCray R. 2005, ApJ, 626, 307 \\

Wilms J., Nowak M.A., Dove J.B., Fender R.P., di Matteo T. 1999, ApJ, 522, 460 \\

Wojdowski P., Clark George W., Levine Alan M., 1998, ApJ, 502, 253 \\

Yentis D., Shulman S., Mckee J.D., Rose W.K., 1977, Astrophys. Lett., 19, 53 \\  
\end{document}